\newcommand{\ignore}[1]{}
\documentstyle[11pt,epsf,sprocl]{article}

\newcommand\be{\begin{equation}}
\newcommand\ee{\end{equation}}
\newcommand\bea{\begin{eqnarray}}
\newcommand\eea{\end{eqnarray}}

\def\bmat{      \left |  \begin{array}{cc} }
\def\emat{ \end{array} \right |    }

\begin{document}

{\hfill Brown-HET-1188}
\vskip30pt

\begin{center}
{\bf Stringy Model for QCD at Finite Density}\\
{\bf and\\Generalized Hagedorn Temperature}
\footnote
{Talk presented by C-I Tan at Workshop on Particle Distributions in Hadronic and Nuclear Collisions. at Univ. of Illinois, Chicago.}

\bigskip

{\bf R. C. Brower}
\\{Physics Department,
        Boston University, Boston, MA 02215}\\
\vskip10pt
       {\bf J. McGreevy, and C-I Tan}\\
Department of Physics, Brown University,
 Providence, RI 02912
\end{center}
\bigskip

\begin{abstract}
Using generic properties of  string theories, we show how interesting
non-perturbative features of QCD can be exploited in  heavy ion collisions. In particular, 
a generalized ``semi-circle" law for the phase diagram  in the temperature-chemical potential
plane is derived.
\end{abstract}

\bigskip

\section{Introduction}

\begin{figure}
\begin{displaymath}
\begin{array}{ccc}
\epsfxsize=5.25cm\epsfbox{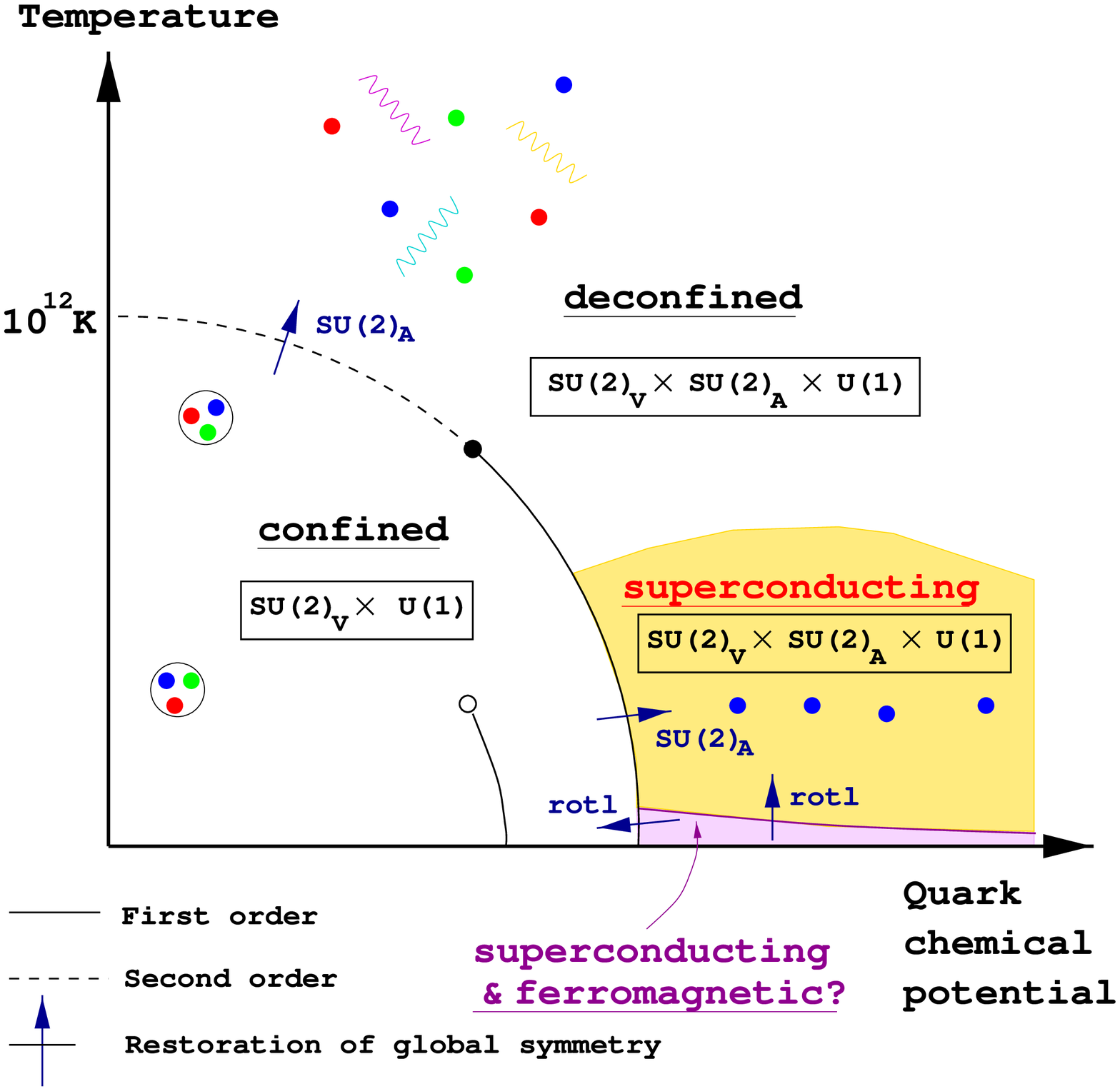}&  &\epsfxsize=5.25cm\epsfbox{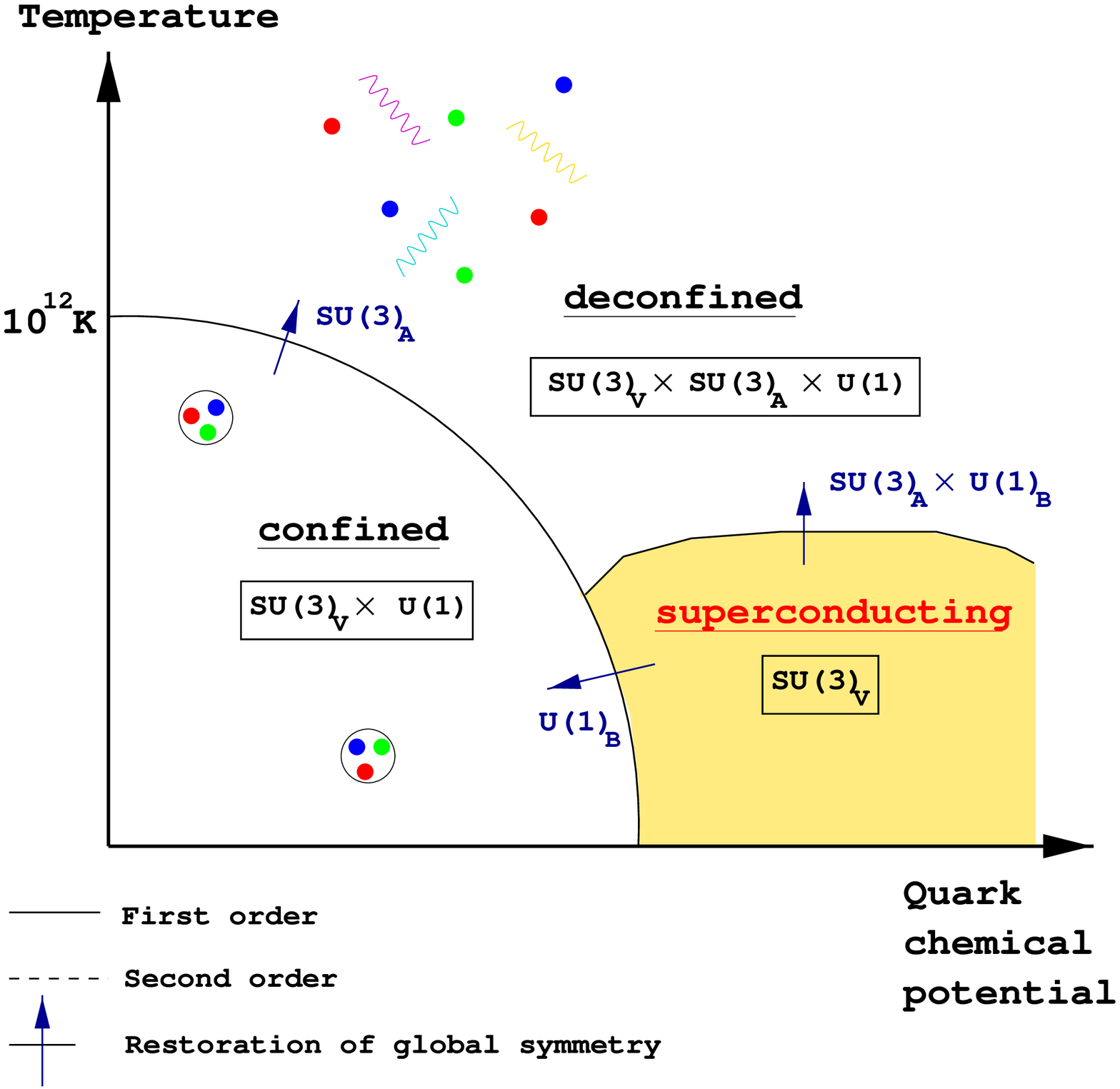}\\
                  (a)               &         &           (b)            
\end{array}
\end{displaymath}
\vspace{.1cm }
\caption{Conjectured phase diagram for QCD: (a) with two massless flavors, (b) with three 
massless flavors.} 
\label{fig:qcd2flv} 

\end{figure}

 It is the belief of many  that
experimental studies of ultra-relativistic nuclear collisions in the near future
should reach  sufficiently  high energy densities so as to reveal the physics of
deconfined quarks and gluons. More recently, the behavior of matter at high baryon
 density has also received much attention. There are growing theoretical indications that there is a 
rich phase
structure for QCD in the temperature-chemical potential, $T-\mu$, plane. For instance, for $SU(2)$ 
flavor, one
believes that the associated chiral-symmetry restoration transition is second order for $\mu$ small, and it turns into first 
order as 
$\mu$ is increased. On the other hand, for $SU(3)$ flavor, the transition is always the first order.
The ``conjectured"  confinement-deconfinement phase diagrams~\cite{MA,SB} for
these two cases are shown in  Fig.~\ref{fig:qcd2flv}(a) and Fig.~\ref{fig:qcd2flv}(b).

Instead of dealing with   features which depend on the specific  
flavor
degrees of freedom, we focus on ``generic features" which can be elucidated by exploiting the 
``stringy" aspect of
QCD spectrum in the confined regime. In particular, we would like to suggest, based on lessons 
 from 
string studies,\cite{DJT1,DJT2,AW,TanParis,StringSM} that interesting non-perturbative features of QCD can  be
learned  at
deconfinement energy densities for non-zero values of chemical potential. In particular,  for heavy ion collisions, we explain how  a stringy representation allows us to derive a
generic expression for the deconfinement curve in the $T-\mu$ plane, which has a similar structure as exhibited in Fig.~\ref{fig:qcd2flv}(a) and Fig.~\ref{fig:qcd2flv}(b).

There are two key ingredients involved:
{\bf Confining force leading to an exponentially growing spectrum,}
and {\bf global symmetries emerging through compactification of extra spatial dimensions.}
We  first
 review how the notion of Hagedorn temperature arises in a stringy theory for
$\mu=0$. We next show how additive quantum numbers naturally arise in string theory through toroidal compactification.
The question of phase diagram in the $T-\mu$ plane will be discussed last.

 \section{Strings at High Energy Densities}

Let us begin by first considering the situation where $\mu=0$. Assuming
thermo-equilibrium can be achieved in hadronic multi-particle production, one
expects that the energy density ${\cal E}$ can be parameterized monotonically in
terms of a temperature, $T$. At low temperature,  ${\cal E}(T)$ can be given
effectively in terms of pion gas. At high temperature and after deconfinement,
one has Stefan-Boltzmann law, (appropriate for gluons and light quarks), which,
for $SU(2)$ flavor, leads to ${\cal E}\simeq 12 T^4$. Indeed, this expectation
has been substantiated by various lattice calculations.~\cite{Satz}
 These ``numerical experiments" 
suggest that the deconfinement density is of the order ${\cal E}_d\simeq 12
T_0^4$, where $T_0\sim 150-200 MeV$.  
We would like to stress that for energy densities  near and/or below ${\cal E}_d$,    the
effective degrees of freedom for QCD are  ``string-like". Because of the confining force, string
excitations can be characterized by rising linear Regge trajectories, leading to an exponentially
growing   particle spectrum. Approaching from the confined phase, deconfinement  is
signaled by a ``Hagedorn temperature",
$T_H\simeq T_0$. Probing this kinematical regime should reveal interesting non-perturbative features
of confined QCD at hadronic scales.  [Atick-Witten have suggested~\cite{AW} that $T_0<T_H$, and we have
previously considered~\cite{TanParis} the possibility that $T_H<T_0$. In what follows, 
we shall assume $T_H\sim T_0$.]

\subsection{Statistical Mechanics of Strings}

For  statistical systems with
a finite number of fundamental degrees of  freedom, it is well-known that a microcanonical and a
canonical descriptions are equivalent.  For strings, this   equivalence breaks down at high energy
densities.  However, it turns out that the canonical partition function  remains useful when
considered as an analytic function of the inverse temperature,
$\beta\equiv 1/T$.

The fundamental quantity in a canonical approach is the partition function:
\be
Z(\beta,V)\equiv Tre^{-\beta  \hat H}
=\sum_{\alpha} e^{-\beta E_{\alpha}},
\ee
where the sum is over all possible multiparticle states of the system.
 For a
microcanonical approach, one works with
 a density function, which counts the number of micro-states,   
$\Omega (E, V)dE
\equiv \sum_{\alpha}{\delta (E - E_{\alpha})}\>\> dE.
$
 Statistical mechanics
based on a microcanonical ensemble is more general, even though it is often more
convenient to work with 
canonical ensemble, {\it e.g.}, when interactions must be included.

 Representing the  Dirac-$\delta$ function by an integral  along 
imaginary axis,
 we find that $
\Omega(E,V)= \sum_{\alpha}
\int_{-i\infty}^{+i\infty} {d\beta\over{2\pi i}}
 \>\> e^{\beta \> (E-E_{\alpha})}.
$
 If one can deform the contour into a region where
interchanging the order of  sum and  integral is allowed, one obtains
\be 
\Omega(E, V)=\int_{\beta_0-i\infty}^{\beta_0+i\infty}
{d\beta\over{2 \pi {i}}}
 \>Z(\beta, V)\>\> e^{\beta E}. 
\label{eq:MC1}
\ee
 The allowed region is labeled by the
interception of the contour with the real axis, $\beta_0$.    One can then
recover canonical partition function from the microcanonical function  via
$Z(\beta,V) =
\int_{0}^{\infty} dE \>\Omega(E, V)\>\> e^{-\beta E},$ which  provides an
alternative analytic definition for $Z(\beta,V)$. 
In order to determine the allowed contour, $\beta_0$, we need to examine {\bf the analytic property
in
$\beta$} for the canonical partition function, $Z(\beta,V)$.

\subsection{Single-Particle Density and Hagedorn Temperature}

Formulating a consistent {\it effective  string theory} for QCD has been one of
the major challenges for string theorists.\cite{StringyQCD} Since we are still far from
accomplishing this task, any
insight into the problem, either theoretical or experimental, can be useful. A common 
feature of all string-like theories is the rapid increase
of mass degeneracies. We shall assume that the desired effective QCD string
theory has an asymptotic  exponential mass degeneracy, which characterizes the
growth of its effective degrees of freedom.  This can be exhibited by calculating the
single-particle density at high energies
$
f(E,V)=V\sum_id(m_i)\int {d^Dp} \>\delta
\left(E-\sqrt{p^2+m_i^2}\right),
$
where $d(m_i)$ is the degeneracy at mass-level $m_i$. More directly, we can work with the inverse
Laplace transform
\be
\tilde f(\beta,V)=V\sum_id(m_i)\int {d^Dp} e^{-\beta \sqrt{p^2+m_i^2}},
\ee
The contour at $\beta_0$ will be determined by the right-most singularity in $\beta$ for $\tilde
f(\beta,V)$. This will be explained in greater detail in the next section, and, in particular, on how
this generalizes in the presence of non-zero chemical potential. 

To be more specific, for an ideal gas of
strings under the Maxwell-Boltzmann (MB) statistics, one has 
$Z(\beta, V)\simeq
e^{\tilde f(\beta, V)}-1.
$
It can be shown that $\tilde
f(\beta, V)$ is analytic for ${\rm Re}\> \beta$ sufficiently large and its
rightmost singularity is  at $\beta=\beta_H>0$, {\it i.e.}, 
\be
\tilde
f(\beta, V)\sim V\left[h(\beta)(\beta-\beta_H)^{\gamma}+\lambda(\beta)\right],
\ee
 where $\gamma>0$ and
both $h(\beta)$ and $\lambda(\beta)$ are analytic around and to the right of $\beta_H$. Given
$Z(\beta,V)$  as an analytic function of $\beta$, the microcanonical function $\Omega(E,V)$ can be
recovered through  Eq. (\ref{eq:MC1}), with
$\beta_0>\beta_H$. That is, {\it  the totality of  physics of microcanonical
approach for free strings has been  encoded in the analyticity of
$Z(\beta,V)$.}~\cite{DJT1,DJT2}  In particular, one finds that
\be
f(E,V)\sim V E^{-(\gamma+1)}e^{\beta_H E}.
\ee
That is, we can characterize a string theory at high energy density minimally  by
 two  exponents, $\gamma$ and $\beta_H$, (the latter is often referred
to as the inverse Hagedorn temperature.) 

We have considered  elsewhere,  in a quantum statistical treatment,  
analytic property of $Z(\beta,V)$ in $\beta$ for various string
theories.~\cite{DJT1,DJT2} We find that, generically, $Z(\beta,V)$ is
 analytic for $Re\>\beta>0$ except at isolated points. For each string
theory, because of the exponential growth in mass degeneracy, there is always
 an isolated rightmost singularity at $\beta=\beta_H$, {\it i.e.}, the
inverse Hagedorn temperature for that theory. There is a finite gap in their
real parts between $\beta_H$ and the next singularity to the left, and this gap
is theory-dependent but calculable.

For conventional systems, Eq. (\ref{eq:MC1}) can often be approximated by a saddle point
 contribution at  $\beta^*$. For strings, at low energies, $\beta^*$ lies to the right of
$\beta_H$.  Under such a condition, the temperature is related to $E$  by
$E=-{\partial \log Z\over \partial \beta}|_{\beta^*}$.
However,
as the  energy density ${\cal E}\equiv E/V$ is
raised, one reaches a point where either the saddle point moves to the left of 
$\beta_H$, or it gets close to $\beta_H$ that the fluctuations about the saddle
point become large. When this occurs, the saddle point approximation to Eq. (\ref{eq:MC1})
 breaks down, and it defines the  density scale ${\cal E}_H$ 
for the region of interest to heavy ion collisions.  Correspondingly, $\beta_H^{-1}=T_H$ will
be identified with the deconfining temperature $T_0$.~\cite{TanParis}

It is worth noting that, for ${\cal E}>{\cal E}_H$, whereas it is no longer meaningful to speak of a
Boltzmann temperature, the statistical mechanics of free strings is  still 
given unambiguously by Eq. (\ref{eq:MC1}). One can in fact push the contour in  (\ref{eq:MC1})  to
the left of the singular point, $\beta_H$, by a finite distance $\eta$,
$\eta>0$. As one moves past this point, one picks up an additional contribution
involving the discontinuity across the cut. Denoting the discontinuity by
$\Delta{Z(\beta,V)}, \beta<\beta_H$,  the large-$E$ behavior of $\Omega(E,V)$ is
dominated by the singularity at $\beta_H$
\be \Omega(E,V)=-\int_{\beta_H-\eta}^{\beta_H}
{d\beta\over{2\pi {i}}}  \Delta{Z(\beta,V)}e^{\beta E}+0(e^{(\beta_H-\eta){E}}),
\hskip 30pt  \eta>0. \ee
  Once $\Delta{Z(\beta,V)}$ is known, the
dominant behavior of $\Omega(E,V)$ can be found. Therefore, the large-$E$ limit of a
free-string theory can best be approached by working first with the canonical
quantity, $Z(\beta,V)$. Applications  to relevant physical systems have been discussed elsewhere~\cite{DJT1,DJT2}.

\section{Counting  Effective Degrees of Freedom at Hadronic
Scales} 

 The basic degrees of freedom of a string theory are  string excitations.
Each excitation can be given a particle attribute, {\it i.e.}, a
mass $m$. The  mass is due to internal string
oscillations; as such, it can take on increasing values, with a corresponding
increase in degeneracies.  Let us illustrate this with the simplest possible bosonic string model.

A classical string is an extended object which ``lives" in space and can be characterized by a set of coordinate 
functions,$
\{ X^{i}(\tau, \sigma)\}, 
$
where $\sigma\epsilon [0, \pi]$ and $i=1,2,\cdots$ specify the set of  spatial directions. Let us 
concentrate on ``closed
strings",  {\it i.e.}, $X(\tau, \sigma+\pi)=X(\tau, \sigma)$.  It follows that the motion can be
represented by a Fourier expansion, 
$X^i=x_0^i + \sum_{n=1}^{\infty}\left( x_n^i\cos 2 n \sigma + \bar x_n^i\sin 2 n \sigma
\right),$ 
where $\{x_0^i\}$ describes the motion of the CM of the string. The set of $\{x_n^i\}$ and $\{\bar
x_n^i\}$  describe amplitudes for  the  internal vibration of the string. Motion for this ``free"
string can be described by an action:
\be
S=\int \frac{d\tau}{4\alpha'}\sum_i\left\{ (\dot x_0^i)^2 + \sum_{n=1} \left[
\left((\dot x_n^i)^2- n^2 (x_n^i)^2\right) + \left( x_n \leftrightarrow \bar x_n
\right)\right]\right\}.
\ee
For the internal  motions, they therefore can also be  described  by an
infinite set of harmonic oscillators, with oscillation frequencies increasing as $n=1, 2, 3,
\cdots.$

One can next turn to quantization.  However, there is the problem of relativistic invariance
which must be taken care of. 
It is of course tempting to maintain covariance all the way. To this end, we need to introduce
$x^{\mu}$, where $\mu = 0, 1, 2, \cdots, D$, where $D$ is the spatial dimension, ({\it e.g.}, for the
real world, $D=3$.) This should be done for both the CM and the internal motions. However, one finds
that not all these variables are independent, and one must  impose proper gauge constraints in
quantization. Or equivalently, one can choose $x^0\equiv \tau$ as the time and work in light-cone
gauge where the independent variables are simply the ``transverse" coordinates: $x_0^i, x_n^i, \bar
x_n^i,$ where $i=1,2,3,\cdots, D-1$.

Classically, one can have separate ``left-moving" and ``right-moving"  vibrations,
{\it i.e.}, independent wave forms of the types $x(\tau+\sigma)$ and $x(\tau-\sigma)$. It is more
convenient to treat these modes separately. For each right-moving (or left-moving) oscillator
mode, we can introduce creation and annihilation operators,  $\tilde
a_n^i$ and $\tilde a_n^{i\dagger}$, or ($a_n^i$ and 
$a_n^{i\dagger}$). The Fock space can then be specified by eigenvalues,
$0,1,2\cdots$, for the corresponding number operators, $ \tilde N_n^i$ and $N_n^i$.

If we denote the Dth direction as ``longitudinal", the dispersion relation for each stationary state 
can be written as that for a particle $
E^2=p_L^2+{\vec p}\>^2 + m^2,
$
$m^2=m_l^2+m_r^2$, where 
\be
m_l^2={1\over {\alpha'}}  \left [-\delta_l+
\sum_{i=1}^{D-1}\sum_{n=1}\left(n
\tilde N_n^i \right)\right],\\ 
m_r^2={1\over {\alpha'}} \left [-\delta_r+\sum_{i=1}^{D-1} \sum_{n=1}\left(n
 N_n^i \right)\right].
\ee
For bosonic string theory, $\delta_l=\delta_r=1$. That is, when one identifies each string
mode as a particle, the  mass originates from the internal string vibrations. One should
also add that translation invariance in
$\sigma$ leads to another constraint
, $m_l^2=m_r^2.$
The details  for the mass spectrum will of course vary from 
theory to theory. However,  generic features for all string models include: (i) equal-spacing
rule in $m^2$ for the mass spectrum, and (ii) exponentially increasing mass degeneracies. To be
specific,  writing
$m^2=\alpha{'}N$, one finds that  the degeneracy factor for a  closed string is given as a
product of left- and right-moving factors,  
$
d(m)=d_l(m_l)d_r(m_r),
$  {\it i.e.},  $d_l(m_l)=\sum_{ 
\tilde N_n^i}^{\infty}\delta\left(N/2+\delta_l-\sum_{i}\sum_{n}n\left(
\tilde  N_n^i\right)\right),
$ and similarly for $d_r(m_r)$. They both  increase  with $\sqrt N$ exponentially.

With these ingredients, one finds that~\cite{OT} 
\be
\tilde f(\beta, V)\sim V\beta \int_E d^2\tau \tau_2^{-(D+3)/2} D_{l}(\bar z) D_{r}(z)
e^{-\beta^2/4\pi
\alpha' \tau_2},
\ee
where the integration in $\tau=\tau_1+i\tau_2$ is over a half-strip region $E: -1/2<\tau_1<1/2,$
$\tau_2>0$. With $\bar z=exp(-2\pi i\bar\tau)$ and $ z=exp(2\pi i\tau)$, $D_{l}$ and $D_{r}$
are generating functions for left- and right-mass degeneracies. By studying the convergence of this
integral  at $\tau_2=0$ as $\beta$ is lowered, the nature of the Hagedorn singularity at
$\beta_H$ can be identified.~\cite{DJT1,OT}

\section{Hagedorn Temperature for $\mu\neq 0$}

We need to first address the question of ``additive" quantum numbers, which
indicate the presence of $U(1)$ global symmetries. This comes most naturally in a string theory by 
a toroidal compactification, which gives rise to  ``winding numbers", $w_i$,  and  ``discrete
momenta", 
$m_i$, one set for each
compactified direction. These discrete quantum numbers are conserved in any process where strings 
are joined or  split. They form the generic representations for conserved charges in a string theory.

If we compactify the jth transverse direction on a
circle of radius $R$, it follows that, for the CM motion, the requirement 
$e^{ip^jx_0^j}=e^{ip^j(x_0^j+2\pi R)}$ leads to a  discrete momentum, $p^j=m_j/R$, $m_j$ an arbitrary
integer. Similarly, the periodicity condition now generalizes to $ x^j(\sigma + \pi)=x^j(\sigma)+
2\pi w_j R$, where $w_j$ is an integral winding number. 
That is, for each compactified direction, $x^j(\tau,\sigma)$, we can quantize the theory
over a ``solitonic" background,
\be
x^j(\tau,\sigma)=\left(\frac{m_j}{R}\right)\>\tau + \left({2 w_jR}\right)\>\sigma.
\ee
This provides an extra  contribution  to the mass term: $(1/2)(m_j/R + w_jR/\alpha')^2$ for $m_l^2$
and
$(1/2)(m_j/R - w_jR/\alpha')^2$ for $m_r^2$ respectively. The constraint,  $m_l^2=m_r^2$,  remains.

\subsection{Statistical Mechanics at Finite Energy and Charge Densities}

In a canonical approach, the partition function becomes
\be
Z(\beta,\mu,V)\equiv Tre^{-\beta  (\hat H+\mu \hat Q)}\equiv Tre^{-\beta  \hat H-\bar \mu \hat Q} 
=\sum_{\alpha} e^{-\beta E_{\alpha}-\bar \mu Q_{\alpha}},
\ee
 where the sum is over all possible multiparticle states of the system.
Note that we have  re-scaled the chemical potential so that $\bar\mu$ is dimensionless. In a 
microcanonical approach, for a system with fixed $E$ and $Q$, one has
$\Omega (E,Q, V)dE
\equiv \sum_{\alpha}\delta_{Q,Q_{\alpha}} {\delta (E - E_{\alpha})}\>\> dE.
$
 Following a similar 
analysis as that for $\mu=0$, 
 we find that 
\be 
\Omega(E,Q, V)=\int_{-i\pi}^{i\pi} {d\bar\mu \over 2  \pi
i}\int_{\beta_0(\bar\mu)-i\infty}^{\beta_0(\bar\mu)+i\infty} {d\beta\over{2 \pi {i}}}
 \>Z(\beta,\mu, V)\>\> e^{\beta E + \bar\mu Q}. 
\label{eq:MC}
\ee
 The allowed region is again labeled by the
interception of the contour with the real axis, $\beta_0(\bar\mu)$.

The analyticity of $Z(\beta,\mu,V)$ can again be determined by the 
single-particle function, which can be symbolically written as 
$
\tilde f(\beta,\bar\mu, V)=\sum_a e^{-\beta \epsilon_a-\bar\mu q_a}$.
Let us be more precise. For each conserved charge, a chemical potential should be introduced. 
For our toroidally compactified strings, $\bar\mu q$ actually represents
$\bar\nu_im_i+\bar\mu_i\omega_i$, where
$m_i$ and
$\omega_i$ are momenta and winding numbers. 

\subsection{Generalized Hagedorn Singularity and Phase Diagram}
We have computed 
\be
\tilde f(\beta, \bar\mu,\bar \nu,V)\sim V\beta \int_E d^2\tau \tau_2^{-(D'+3)/2}
D_{l}(\bar z) D_{r}(z) e^{(-\beta^2/4\pi
\alpha' \tau_2)} W(\bar z, z),
\ee
where the additional factor due to compactification is
\be
W(\bar z,  z)=\Pi_{i}^{'}\sum_{m_i,w_i}{\bar z}^{{1\over 2}\left(\frac{m_i}{ \bar R} + {w_i \bar R}
\right)^2}{
z}^{{1\over 2}\left(\frac{m_i}{\bar R} - {w_i \bar R}\right)^2}e^{-\left(\bar\nu_i m_i +\bar\mu_i
w_i\right)},
   \ee
where $\bar R\equiv \alpha'^{-1/2} R$, and the product is over the compactified directions. Again, by examining the divergence
at
$\tau_2\rightarrow 0$ as one lowers $\beta$, one finds that the generalized Hagedorn temperature is
given by
\be
\frac{2\beta_H(\mu,\nu)}{{\left(\alpha'\right)}^{\frac{1}{2}}}=\left [\bar\omega_l +
\sum_i\left (\bar\nu_i\bar R + {\bar\mu_i\over \bar R}\right)^2\right]^{1/2}+\left [\bar\omega_r
+\sum_i\left (\bar\nu_i\bar R - {\bar\mu_i\over
\bar R}\right)^2\right]^{1/2}.
\label{eq:TMu}
\ee
In particular, $\beta_H(0,0)=(2\pi^2 \alpha')^{\frac{1}{2}}\left(\sqrt {\omega_l} +\sqrt
{\omega_r}\right) =\beta_H$, ($\bar\omega_{l,r}=8\pi^2\omega_{l,r}$), is simply the inverse Hagedorn 
temperature identified previously. For standard string theories, $\omega_{l,r}$ take on values 1 or 2.

We are now in the position to apply our result to the case of QCD at finite density. 
Here, we are interested in the phase diagram for the deconfinement transition in the $T-\mu$ plane.
We therefore consider the simplest possible situation where we set  all
$\mu_i$ and
$\nu_i$ to be zero, except one, which will be referred to as the ``baryonic chemical potential".

In order for us to make use the above result, Eq. (\ref{eq:TMu}), we
must recognize that
$\bar\mu=\beta\mu$. (Alternatively, we could have worked directly with $\mu$ without involving
the reduced chemical potential $\bar\mu$. We have explicitly verified that the same result would be
reached. In particular, in the large volume limit, the $\bar\mu$-integration in Eq. (\ref{eq:MC}) is
dominated by a saddle point with $\mu$ real.~\cite{BMT}) Let us denote the phase boundary by
$\beta_H(\mu)\equiv T(\mu)^{-1}$. It follows that 
\be
\frac{2}{\sqrt \alpha'}=\left[\bar\omega_lT(\mu)^2 +\left(\frac{\mu}{\bar R}
\right)^2\right]^{1/2}+\left[\bar\omega_rT(\mu)^2 +\left(\frac{\mu}{\bar R}
\right)^2\right]^{1/2},
\ee
which corresponds to  a ``distorted" semi-circle law. To be explicit, consider  the case
$\bar\omega_l=\bar\omega_r$. One finds that 
$\left({T(\mu)\over T_H}\right)^2+ \left(\frac{\mu}{\mu_0}\right)^2=1.
$
Note that $\mu_0=\alpha'^{-1}R$, which is precisely the mass for one unit of 
baryonic charge.  This result can  be
expressed more simply as
\be
T(\mu)=T_H\sqrt{1-\frac{\mu^2}{\mu_0^2}}\>\>,
\ee
as depicted in Fig.~\ref{fig:qcd2flv}(a) and (b). 

\section{Remarks}
It is well understood that the character of  QCD changes depending on  the
nature of available  probes. At short distances, the basic degrees of freedom
are quarks and gluons.  As one moves to larger distance scales, the QCD
coupling increases and one  enters the non-perturbative
regime.  Short of resulting to lattice Monte Carlo studies, the most promising
tool for a non-perturbative treatment of QCD  which builds in naturally
quark-gluon confinement remains the large-$N$ expansion. In this approach,
although the vacuum of QCD at hadronic scales is complicated, model studies
suggest that  the effective degrees of freedom of QCD  can most profitably
be expressed in terms of ``extended objects". Indeed, low-lying hadron spectrum
suggests that they can be understood as ``string excitations".
  In high-energy soft
hadronic collisions~\cite{DPM} where the interactions are mostly peripheral, it is
possible to ``see" the dominant string excitations in terms of the exchanges
of high-lying Regge trajectories having a closed string color topology.

Here we have concentrated on deriving the phase-boundary for the deconfining transition at 
non-zero baryon chemical potential. More interestingly, our approach can provide a framework for
exploring interesting non-perturbative features of QCD at deconfinement energies for non-zero baryon
densities through our microcanonical approach, Eq. (\ref{eq:MC}). This will be discussed in a
separate publication.~\cite{BMT}

\end{document}